\input harvmac
\input epsf
\input tables

\def\gsim{{~\raise.15em\hbox{$>$}\kern-.85em
          \lower.35em\hbox{$\sim$}~}}
\def\lsim{{~\raise.15em\hbox{$<$}\kern-.85em
          \lower.35em\hbox{$\sim$}~}} 
\def\Re{{\cal R}e}
\def\Im{{\cal I}m}
\def\epsK{{\varepsilon}_K}
\def\epspK{{\varepsilon}^\prime_K}
\def\epe{\varepsilon^\prime/\varepsilon}

\def\O{{\cal O}}

\def\YGTitle#1#2{\nopagenumbers\abstractfont\hsize=\hstitle\rightline{#1}%
\vskip .4in\centerline{\titlefont #2}\abstractfont\vskip .3in\pageno=0}
\YGTitle{WIS-99/29/Aug-DPP, SISSA-95/99/EP, TUM-HEP-354/99, hep-ph/9908382}
{\vbox{
\centerline{Probing Supersymmetric Flavor Models with $\epe$}}}
\bigskip
\centerline{G. Eyal$^a$, A. Masiero$^{b,c}$, Y. Nir$^a$ and L. Silvestrini$^d$}
\smallskip
\centerline{\it $^a$ Department of Particle Physics}
\centerline{\it Weizmann Institute of Science, Rehovot 76100, Israel}
\centerline{\it $^b$ SISSA-ISAS, Via Beirut 2-4, Trieste, Italy}
\centerline{\it $^c$ INFN, Sezione di Trieste, Trieste, Italy}
\centerline{\it $^d$ Physik Department, Technische Universit\"at M\"unchen}
\centerline{\it D-85748 Garching, Germany}
\bigskip

\baselineskip 18pt
\noindent
We discuss the supersymmetric contribution to $\epe$ in various
supersymmetric flavor models. We find that in alignment models
the supersymmetric contribution could be significant while in
heavy squark models it is expected to be small. The situation is
particularly interesting in models that solve the flavor problems 
by either of the above mechanisms and the remaining CP problems
by means of approximate CP, that is, all CP violating phases are small. 
In such models, the standard model contributions 
cannot account for $\epe$ and a failure of the supersymmetric contributions
to do so would exclude the model. In models of alignment and approximate CP,
the supersymmetric contributions can account for $\epe$ only
if both the supersymmetric model parameters and the hadronic parameters
assume rather extreme values. Such models are then strongly disfavored by the
$\epe$ measurements. Models of heavy squarks and approximate CP are excluded.  
\bigskip
 
\baselineskip 18pt
\leftskip=0cm\rightskip=0cm
 
\Date{}

\newsec{Introduction}
The $\epe$ parameter, signifying direct CP violation, has now been
measured with impressive accuracy 
\nref\NAto{H. Burkhardt {\it et al.}, NA31 collaboration, 
 Phys. Lett. B206 (1988) 169.}%
\nref\NAton{G.D. Barr {\it et al.}, NA31 collaboration,
 Phys. Lett. B317 (1993) 233.}%
\nref\Esto{L.K. Gibbons {\it et al.}, E731 collaboration,
 Phys. Rev. Lett. 70 (1993) 1203.}%
\nref\KTeV{A. Alavi-Harati {\it et al.}, KTeV collaboration,
 Phys. Rev. Lett. 83 (1999) 22, hep-ex/9905060.}%
\nref\NAfe{V. Fanti {\it et al.}, NA48 collaboration, hep-ex/9909022.}%
\refs{\NAto-\NAfe}:
\eqn\expepe{\Re(\epe)=(2.12\pm0.46)\times10^{-3}.}
The theoretical interpretation of this result suffers from large
hadronic uncertainties. Within the Standard Model, the theoretically preferred
range is somewhat lower than the experimental range of eq. \expepe\
\nref\CFMR{M. Ciuchini, E. Franco, G. Martinelli and L. Reina, 
 Phys.Lett. B301 (1993) 263, hep-ph/9212203.}%
\nref\BJL{A.J. Buras, M. Jamin and M.E. Lautenbacher,
 Nucl.Phys. B408 (1993) 209, hep-ph/9303284;
 Phys. Lett. B389 (1996) 749, hep-ph/9608365.}%
\nref\CFMRS{M. Ciuchini, E. Franco, G. Martinelli, L. Reina and L. Silvestrini,
 Z. Phys. C68 (1995) 239, hep-ph/9501265.}%
\nref\Ciuc{M. Ciuchini,
 Nucl. Phys. Proc. Suppl. 59 (1997) 149, hep-ph/9701278.}%
\nref\BEFL{S. Bertolini, J.O. Eeg, M. Fabbrichesi and E.I. Lashin,
 Nucl. Phys. B514 (1998) 93, hep-ph/9706260.}
\nref\Burepe{A.J. Buras, hep-ph/9806471.}%
\nref\KNS{Y.-Y. Keum, U. Nierste and A.I. Sanda,
 Phys. Lett. B457 (1999) 157, hep-ph/9903230.}%
\nref\BBGJ{S. Bosch {\it et al.}, hep-ph/9904408.}%
\nref\HKPS{T. Hambye, G.O. Koehler, E.A. Paschos and P.H. Soldan, 
 hep-ph/9906434.}%
\nref\BuSi{A.J. Buras and L. Silvestrini,
 Nucl.Phys. B546 (1999) 299, hep-ph/9811471.}%
(for recent work, see \refs{\CFMR-\BuSi}\ and references
therein). Yet, if all the hadronic parameters are taking values
at the extreme of their reasonable ranges, the experimental result
can be accommodated. 

While \expepe\ does not provide unambiguous evidence for new physics, 
it is still useful in testing extensions of the Standard Model.
Models where $\epspK$ is suppressed and/or $\epsK$ enhanced are
disfavored. Models that allow significant new contributions to
$\epe$ may be favored if future improvement in the experimental
measurement and, in particular, in the theoretical calculation will 
prove that the Standard Model fails to account for its large value.
Investigations of the supersymmetric contributions to $\epe$
in view of the recent experimental results have been presented in refs.
\nref\MaMu{A. Masiero and H. Murayama, hep-ph/9903363.}%
\nref\BDM{K.S. Babu, B. Dutta and R.N. Mohapatra, hep-ph/9905464.}%
\nref\KhKo{S. Khalil and T. Kobayashi, hep-ph/9906374.}%
\nref\BJKP{S. Baek, J.-H. Jang, P. Ko and J.H. Park, hep-ph/9907572.}%
\nref\BCS{R. Barbieri, R. Contino and A. Strumia, hep-ph/9908255.}%
\refs{\MaMu-\BCS}.
Most interestingly, in models where CP is an approximate symmetry
of electroweak interactions, that is, all CP violating phases are 
small, it is clear already at present that the Standard Model cannot 
explain \expepe. These models should then provide new contributions to
fully account for $\epe$. A failure to do so would mean that the
model is excluded. In this work we examine two such classes of 
supersymmetric flavor models. In the first class of models, a horizontal 
Abelian symmetry solves the supersymmetric flavor problems by means
of alignment and approximate CP solves the remaining CP problems.
In the second class, the first two sfermion generations are heavy,
thus solving most flavor problems, while mild alignment and
approximate CP solve the remaining flavor and CP problems.

The plan of this paper goes as follows. In section 2, we present the
possible supersymmetric sources of $\epe$ and discuss their uncertainties.
In sections 3-5, we estimate the various contributions in models of
Abelian horizontal symmetries. In section 3, we study models where
a horizontal symmetry explains the Yukawa hierarchy. In section 4, we
require that, in addition, the symmetry solves the supersymmetric
flavor problems. In section 5, we add the assumption of approximate CP.
Models with heavy first two squark generations are discussed in section 6. 
We summarize our results in section 7.

\newsec{Supersymmetric contributions to $\epe$}
In generic supersymmetric models, there are potentially
many new contributions to $\epe$ from loop diagrams involving
intermediate squarks and gluinos, charginos or neutralinos.
If there is some degeneracy between squarks, then a convenient way 
to parametrize these contributions is by using the $(\delta_{MN}^q)_{ij}$ 
parameters. In the basis where quark masses and gluino couplings are diagonal, 
the dimensionless $(\delta_{MN}^q)_{ij}$ parameters
stand for the ratio between $(M^2_{\tilde q})^{MN}_{ij}$, the $(ij)$ 
entry ($i,j=1,2,3$) in the mass-squared matrix for squarks ($M,N=L,R$ 
and $q=u,d$), and $\tilde m^2$, the average squark mass-squared.
If there is no mass degeneracy among squarks, then these parameters
can be related to the supersymmetric mixing angles. Defining $K_L^d$ 
($K_R^d$) to be the mixing matrix between left-handed (right-handed) down 
quarks and the scalar partners of left-handed (right-handed) down quarks,
we have, {\it e.g.}, $(\delta_{LL}^d)_{12}\sim(K_L^d)_{12}$. 

For supersymmetry to account for $\epe$, at least one of the following 
conditions should be met
\nref\GMS{E. Gabrielli, A. Masiero and L. Silvestrini,
 Phys. Lett. B374 (1996) 80, hep-ph/9509379.}%
\nref\GGMS{F. Gabbiani, E. Gabrielli, A. Masiero and L. Silvestrini,
 Nucl. Phys. B477 (1996) 321, hep-ph/9604387.}%
\nref\CoIs{G. Colangelo and G. Isidori, JHEP 9809 (1998) 009, hep-ph/9808487.}%
\nref\LuSi{L. Silvestrini, hep-ph/9906202.}%
\refs{\BuSi-\LuSi}:
\eqn\susyepe{\eqalign{
\Im[(\delta^d_{LL})_{12}]&\sim\lambda\left({\tilde m\over500\ GeV}\right)^2,\cr
\Im[(\delta^d_{LR})_{12}]&\sim\lambda^7\left({\tilde m\over500\ GeV}\right),\cr
\Im[(\delta^d_{LR})_{21}]&\sim\lambda^7\left({\tilde m\over500\ GeV}\right),
\cr}}
\eqn\Zdsepe{\eqalign{
\Im[(\delta^u_{LR})_{13}(\delta^u_{LR})_{23}^*]&\sim\lambda^2,\cr
\Im[V_{td}(\delta^u_{LR})_{23}^*]&\sim\lambda^3\left({M_2\over m_W}\right),\cr
\Im[V_{ts}^*(\delta^u_{LR})_{13}]&\sim\lambda^3\left({M_2\over m_W}\right).\cr}}
Here $\lambda=0.2$ is a small parameter of order of the Cabibbo angle
that is convenient to use in the context of flavor models.

Let us first discuss the three options of eq. \susyepe. The first of these 
conditions violates constraints from $\Delta m_K$ and $\epsK$. 
Therefore, independent of the supersymmetric flavor model,
it cannot be satisfied. On the other hand, the requirements on
$\Im[(\delta^d_{LR})_{12}]$ or $\Im[(\delta^d_{LR})_{21}]$ pose
no phenomenological problem. Moreover, we will see in the next section
that such values are not impossible within our theoretical framework. 
We therefore investigate more carefully the uncertainties in the corresponding 
condition. Using the following expression for the matrix element of the
chromomagnetic operator 
\ref\BEF{S. Bertolini, J.O. Eeg and M. Fabbrichesi,
 Nucl. Phys. B449 (1995) 197, hep-ph/9409437.}:
\eqn\eqme{
\vev{(\pi\pi)_{I=0}|Q_g|K^0} = \sqrt{{3\over2}}{11\over16\pi^2}
{\vev{\bar q q}\over F_\pi^3}m_\pi^2 B_G,}
the expression for the Wilson coefficients at the SUSY scale given in
\GGMS\ and LO QCD corrections, one can write the contribution of $\Im
[(\delta^d_{LR})_{12}]$ and $\Im [(\delta^d_{LR})_{21}]$ to $\epe$ as follows 
\ref\BCIRS{A.J. Buras, G. Colangelo, G. Isidori, A. Romanino and
 L. Silvestrini, hep-ph/9908371.}: 
\eqn\LRdot{\eqalign{|\epe|\ =\ &58\ B_G\left[
{\alpha_s(m_{\tilde g})\over\alpha_s(500\ GeV)}\right]^{23/21}\left(
{158\ MeV\over m_s(m_c)+m_d(m_c)}\right)\cr \times&\left(
{500\ GeV\over m_{\tilde g}}\right)
\left|\Im\left[(\delta^d_{LR})_{12}-(\delta^d_{LR})^*_{21}\right]\right|.\cr}}
Here the parameter $B_G$ accounts for possible deviations of the hadronic
matrix element in eq. \eqme\ from the value obtained at lowest
order in the chiral quark model. Given the large uncertainties from higher
order contributions and the ``anomalous'' $m_\pi^2$ suppression of the
matrix element in eq. \eqme, we use the conservative range $B_G\leq5$
(see ref. \BCIRS\ for a detailed derivation of eq. \LRdot\ and of the related
hadronic uncertainties). Using 
\eqn\lowerpar{\eqalign{
\epe\ \gsim&\ 1\times10^{-3},\cr
m_{\tilde g}\ \gsim&\ 150\ GeV,\cr
m_s(m_c)\ \gsim&\ 110\ MeV,\cr}}
we get a lower bound:
\eqn\lowerLRot{\Im(\delta^d_{LR})_{12}\gsim7\times10^{-7},}
that is $\O(\lambda^9)$ or even $\O(\lambda^{10})$ if $\lambda\sim0.24$.
A similar bound applies to $\Im[(\delta^d_{LR})_{21}]$.

We now turn to the three options in eq. \Zdsepe. These contributions
to $\epe$ arise by inducing an effective $Z_{ds}$ coupling, where \BuSi:
\eqn\defZds{{\cal L}_{\rm FC}^Z={G_F\over\sqrt{2}}{em_Z^2\over2\pi^2}
{\cos\theta_W\over\sin\theta_W}Z_{ds}\bar s\gamma_\mu(1-\gamma_5)dZ^\mu
+{\rm h.c.}.}
The contribution of such an effective coupling to $\epe$ is given by
\eqn\epeZds{\epe=\Im Z_{ds}\left[1.2-\left(
{158\ MeV\over m_s(m_c)+m_d(m_c)}\right)^2\left|r_Z^{(8)}\right|
B_8^{(3/2)}\right],}
where $B_8^{(3/2)}$ is the non-perturbative parameter describing
the hadronic matrix element of the electroweak penguin operator and
$\left|r_Z^{(8)}\right|$ is a calculable renormalization scheme
independent parameter. We consider the following ranges \BuSi:
\eqn\ranhapa{\eqalign{6.5\leq\left|r_Z^{(8)}\right|\leq&\ 8.5,\cr
0.6\leq B_8^{(3/2)}\leq&\ 1.0.\cr}}
Then, for $\Im Z_{ds}$ to account for $\epe$, we need
\eqn\epeonZds{-\Im Z_{ds}\gsim{\epe\over16}.}
To find the lower bound on $\Im[V_{td}(\delta^u_{LR})_{23}^*]$,
we have performed a more careful analysis of its relation to $Z_{ds}$.
We scanned the following range of supersymmetric parameters:
\eqn\suspar{\eqalign{
-300\ GeV\leq\mu&\leq300\ GeV,\cr 
100\ GeV\leq M_2&\leq250\ GeV,\cr 
3M_2\leq m_{\tilde{Q}}&\leq5M_2,\cr 
0.4m_{\tilde{Q}}\leq m_{\tilde{t}_R}&\leq m_{\tilde{Q}},\cr}}
and discarded points in which chargino masses are $\leq90\ GeV$. 
We found that  
\eqn\ZdsVdel{\Im Z_{ds}\leq0.03\ \Im[V_{td}(\delta^u_{LR})_{23}^*].}
Together with eq. \epeonZds\ and the lower bound on $\epe$ quoted
in \lowerpar, we find then the following lower bound:
\eqn\weakZds{\Im[V_{td}(\delta^u_{LR})_{23}^*]\gsim2\times10^{-3},}
that is $\O(\lambda^4)$. Bounds similar to \ZdsVdel\ and \weakZds\
apply to $\Im[V_{ts}^*(\delta^u_{LR})_{13}]$. The lower bound on
$\Im[(\delta^u_{LR})_{13}(\delta^u_{LR})_{23}^*]$ is stronger, of 
${\cal O}(\lambda^3)$.

\newsec{Abelian Horizontal Symmetries}
Models of Abelian horizontal symmetries are able to provide a
natural explanation for the hierarchy in the quark and lepton
flavor parameters
\ref\FrNi{C.D. Froggatt and H.B. Nielsen, Nucl. Phys. B147 (1979) 277.}.
The symmetry is broken by a small parameter $\lambda$ which is usually
taken to be of the order of the Cabibbo angle, $\lambda\sim0.2$.
The hierarchy in the flavor parameters is then a result of the
selection rules related to the approximate horizontal symmetry.
In the supersymmetric framework, holomorphy also plays a role in
determining the Yukawa parameters
\ref\LNS{M. Leurer, Y. Nir and N. Seiberg,
 Nucl. Phys. B398 (1993) 319, hep-ph/9212278.}.

For the sake of definiteness, we take the following order of magnitude 
estimates for the various quark mass ratios and mixing angles:
\eqn\phypar{\eqalign{
m_{u}/m_{c}\sim\lambda^{3},\ \ \  &m_{c}/m_{t}\sim\lambda^{4},\ \ \ 
m_{t}/\vev{\phi_{u}}\sim 1,\cr
m_{d}/m_{s}\sim \lambda^{2},\ \ \  &m_{s}/m_{b}\sim\lambda^{2},\ \ \ 
m_{b}/m_{t}\sim\lambda^{3},\cr
|V_{us}|\sim \lambda,\ \ \ &|V_{cb}|\sim \lambda^{2},\ \ \ 
|V_{ub}|\sim \lambda^{3}.\cr}}
Within models of a single horizontal $U(1)$ symmetry, this set of parameters 
determines all the horizontal charges, leading to the following structure
of the quark mass matrices:
\eqn\qmama{M_u\sim\vev{\phi_u}\pmatrix{\lambda^7&\lambda^5&\lambda^3\cr
\lambda^6&\lambda^4&\lambda^2\cr \lambda^4&\lambda^2&1\cr},\ \ \ 
M_d\sim\vev{\phi_d}\lambda^3\tan\beta\pmatrix{\lambda^4&\lambda^3&\lambda^3\cr
\lambda^3&\lambda^2&\lambda^2\cr \lambda&1&1\cr}.}
A similar hierarchy appears also in the $(LR)$ blocks of the corresponding
squark mass-squared matrices:
\eqn\sqLR{(M^2_{\tilde u})^{LR}_{ij}\sim \tilde m(M_u)_{ij},\ \ \ 
(M^2_{\tilde d})^{LR}_{ij}\sim \tilde m(M_d)_{ij}.}
We emphasize that eq. \sqLR\ does {\it not} imply
$(M^2_{\tilde q})^{LR}\propto M_q$. The coefficients 
of order one are independent and different in the respective entries. Only 
the parametric suppression is the same for the squarks and the quarks.

Eqs. \qmama\ and \sqLR\ allow us to estimate the values of the $\delta_{LR}$
parameters of eq. \susyepe\ and \Zdsepe. We get:
\eqn\houoepe{\eqalign{
(\delta^d_{LR})_{12}&\sim{m_s|V_{us}|\over\tilde m}\sim\lambda^6\
{m_t\over\tilde m},\cr
(\delta^d_{LR})_{21}&\sim{m_d\over|V_{us}|\ \tilde m}\sim\lambda^6\
{m_t\over\tilde m},\cr
(\delta^u_{LR})_{13}&\sim{m_t|V_{ub}|\over\tilde m}\sim\lambda^3\
{m_t\over\tilde m},\cr
(\delta^u_{LR})_{23}&\sim{m_t|V_{cb}|\over\tilde m}\sim\lambda^2\
{m_t\over\tilde m}.\cr}}

Taking into account that $|V_{td}|\sim\lambda^3$ and $|V_{ts}|\sim\lambda^2$,
we learn that the three options in eq. \Zdsepe\ are of order
$\lambda^{5-7}$. We compare this to the requirements given in eq. \weakZds\
(and the discussion below this equation) and conclude that, in models
of Abelian horizontal symmetries, the contributions to $Z_{ds}$ involving
$\tilde t_R$ cannot account for $\epe$.

On the other hand, $(\delta^d_{LR})_{12}$ and $(\delta^d_{LR})_{21}$ are 
large enough to allow for a supersymmetric explanation of $\epe$ \MaMu.
This is the case even if the supersymmetric scale is not particularly low,
say $\tilde m\sim500\ GeV$. If the supersymmetric scale is lower,
then a supersymmetric explanation of $\epe$ is possible even for small
phases, $\phi_{CP}\sim\lambda^{2-4}$, if other relevant 
parameters take extreme values, as in eq. \lowerLRot. 

\newsec{Alignment}
It is possible to solve the supersymmetric flavor problems by the
mechanism of alignment
\nref\NiSe{Y. Nir and N. Seiberg,
 Phys. Lett. B309 (1993) 337, hep-ph/9304307.}%
\nref\LNSb{M. Leurer, Y. Nir and N. Seiberg,
 Nucl. Phys. B420 (1994) 468, hep-ph/9310320.}%
\nref\GrNiL{Y. Grossman and Y. Nir,
 Nucl. Phys. B448 (1994) 30, hep-ph/9502418.}%
\nref\RaNi{Y. Nir and R. Rattazzi,
 Phys. Lett. B382 (1996) 363, hep-ph/9603233.}%
\refs{\NiSe-\RaNi}, whereby the mixing matrices for gaugino couplings 
have very small mixing angles. Alignment arises naturally in the
framework of Abelian horizontal symmetries. Simple models give 
supersymmetric mixing angles that are similar to the corresponding
CKM mixing angles. However, for the mixing between the first two
down squark generations, a much more precise alignment is needed,
that is,
\eqn\Kdot{(K_L^d)_{12}\lsim\lambda^2,\ \ \ (K_R^d)_{12}\lsim\lambda^2,\ \ \ 
(K_L^d)_{12}(K_R^d)_{12}\lsim\lambda^6.}

Eq. \Kdot\ implies that some of the entries in $M_d$ should be suppressed
compared to their `naive' values, given in eq. \qmama. In particular, the
following constraints should be satisfied:
\eqn\alignMd{\eqalign{
(M_d)_{12}/(M_d)_{22}\ \lsim&\ \lambda^2,\cr
(M_d)_{21}/(M_d)_{22}\ \lsim&\ \lambda^2,\cr
(M_d)_{12}(M_d)_{21}/[(M_d)_{22}]^2\ \lsim&\ \lambda^6.\cr}}
Additional constraints apply to $(M_d)_{31}$ and $(M_d)_{32}$, but
they are irrelevant to our study here.

To achieve the required suppression, one has to employ a more complicated
Abelian horizontal symmetry. The models of refs. \refs{\NiSe-\RaNi}\ use
$U(1)\times U(1)$ symmetries. Then, it is possible to retain all the
`good' predictions of eq. \phypar\ and, at the same time, have
the Yukawa couplings that are relevant to \Kdot\ vanish due to
holomorphy of the superpotential, that is, $(M_d)_{12}=0$, $(M_d)_{21}=0$, 
either $(M_d)_{13}$ or $(M_d)_{32}=0$ and either $(M_d)_{31}$ or 
$(M_d)_{23}=0$. The vanishing of these entries is exact
in the basis where the horizontal charges are well defined.
In this basis the kinetic terms are not canonically normalized.
When we transform to a basis with canonical normalization of the
kinetic terms, the holomorphic zeros are lifted \LNSb. However, they
are suppressed by at least a factor of $\lambda^2$ compared to their
naive values of eq. \qmama\
\nref\Eyal{G. Eyal and Y. Nir, Nucl. Phys. B528 (1998) 21, hep-ph/9801411.}%
\refs{\LNSb,\Eyal}. The suppression could be much stronger but
it is always an even power of the breaking parameter.

The entries in the $LR$-block of the down-squark mass-squared matrix,
that is $(M^2_{\tilde d})^{LR}_{ij}$, are suppressed in a similar way to
the corresponding entries in the down quark mass matrix, $(M_d)_{ij}$.
Consequently, the alignment requirements \Kdot\ affect directly 
$(M^2_{\tilde d})^{LR}_{12}$ and $(M^2_{\tilde d})^{LR}_{21}$ that
are relevant to $\epe$. Independent of the details of the model,
we find that in the framework of alignment, we have
\eqn\aligepe{\eqalign{
(\delta^d_{LR})_{12}&\lsim{m_s|V_{us}|\over\tilde m}\lambda^2\sim\lambda^8\
{m_t\over\tilde m},\cr
(\delta^d_{LR})_{21}&\lsim{m_d\over|V_{us}|\ \tilde m}\lambda^2\sim\lambda^8\
{m_t\over\tilde m}.\cr}}
The values in eq. \aligepe\ should be compared with the phenomenological input 
of eq. \susyepe. It is interesting that for central values of the hadronic
parameters, the supersymmetric contributions to $\epe$ in models of
alignment can naturally be of the required order of magnitude.
For this to happen, the models have to satisfy two conditions:
\item{(i)} The alignment has to be minimal in the sense explained above,
that is either $|(K_L^d)_{12}|\sim\lambda^3$ or 
$|(K_R^d)_{12}|\sim\lambda^3$ should hold.
\item{(ii)} The relevant phase is of order one.

We note that a situation where both $|(K_L^d)_{12}|$ and $|(K_R^d)_{12}|$
are of order $\lambda^3$ and with a phase of order one is forbidden
since it will give a contribution to $\epsK$ that is too large.

\newsec{Approximate CP}
The requirement that supersymmetric contributions to flavor changing 
CP violation, that is the $\epsK$ parameter, and to flavor diagonal
CP violation, such as the electric dipole moment of the neutron $d_N$,
are not too large, puts severe constraints on the supersymmetric parameters
(for a review, see
\ref\GNR{Y. Grossman, Y. Nir and R. Rattazzi, in {\it Heavy Flavours II}, 
eds. A.J. Buras and M. Lindner (World Scientific), hep-ph/9701231.}). 
For example, the $\epsK$ constraints read
\eqn\epsKcon{\eqalign{
\sqrt{\Im[(\delta^d_{LL})^2_{12}]}&\lsim\lambda^3,\cr
\sqrt{\Im[(\delta^d_{RR})^2_{12}]}&\lsim\lambda^3,\cr
\sqrt{\Im[(\delta^d_{LL})_{12}(\delta^d_{RR})^2_{12}]}&\lsim\lambda^5.\cr}}
The third constraint is particularly strong. For example, for 
$|(\delta^d_{LL})_{12}|\sim|(\delta^d_{RR})_{12}|\sim\lambda^3$, the
phase in their product needs to be smaller than $\O(\lambda^4)$.
The $d_N$ constraint generically requires that the flavor diagonal phases 
$\phi_A$ (related to the trilinear scalar couplings)
and $\phi_B$ (related to bilinear $\phi_u\phi_d$ terms) are small,
\eqn\dNcon{\phi_A\lsim10^{-2},\ \ \ \phi_B\lsim10^{-2}.}
Approximate CP is then a well motivated option in supersymmetric models.

Before discussing models with alignment and approximate CP, we would
like to make the following comments. It is possible to construct models 
of alignment with the following features \RaNi:
\item{(i)} The alignment is precise enough that it solves not only the
supersymmetric flavor problems but also the supersymmetric $\epsK$ problem.
\item{(ii)} The $\phi_A$ and $\phi_B$ phases are small enough to solve the
supersymmetric $d_N$ problem.
\item{(iii)} The Kobayashi-Maskawa phase is of order one, so that
$\epe$ can be explained by Standard Model contributions.

The explicit models of ref. \RaNi\ where these features are realized
are very constrained and almost unique. We learn that, on one hand, 
approximate CP is not a necessary ingredient in models of alignment but, 
on the other hand, it allows simpler and less constrained models of this type.

We focus then on models where all flavor problems are solved by
alignment, but the CP problems are solved by approximate CP.
The main point is that, independent of the details of the model,
the CP violating phases in this framework are suppressed by even
powers of the breaking parameter. The reason for that is as follows.
In the framework of Abelian horizontal symmetries, all contributions
to a given term should carry the appropriate horizontal charge. If the 
CP violating
contribution does not appear in leading order, then it should be suppressed
by a neutral combination of the breaking parameter, such as $\lambda\lambda^*$.
An odd power of the breaking parameter always carries a horizontal charge.

The conclusion is that, in models of approximate CP, the imaginary
part of any $(\delta^q_{MN})_{ij}$ term is suppressed by, at least,
a factor of $\lambda^2$ compared to the real part. In particular, we have
\eqn\apprepe{\eqalign{\Im(\delta^d_{LR})_{12}&
\lsim{m_s|V_{us}|\over\tilde m}\lambda^4\sim\lambda^{10}{m_t\over\tilde m},\cr
\Im(\delta^d_{LR})_{21}&\lsim{m_d\over|V_{us}|\ \tilde m}\lambda^4\sim
\lambda^{10}{m_t\over\tilde m}.\cr}}
These are rather low values. They are consistent with the experimental
constraint of eq. \lowerLRot\ only if all the following conditions are
simultaneously satisfied:
\item{(i)} The suppression of the relevant CP violating phases is `minimal', 
that is $\O(\lambda^2)$.
\item{(ii)} The alignment of the first two down squark generations is 
`minimal', that is $\O(\lambda^2)$.
\item{(iii)} The mass scale for the supersymmetric particles is low, 
$\tilde m\sim150\ GeV$.
\item{(iv)} The hadronic matrix element is larger than what hadronic
models suggest, $B_G\sim5$.
\item{(v)} The mass of the strange quark is at the lower side of the
theoretically preferred range, $m_s(m_c)\sim110\ MeV$.
\item{(vi)} The value of $\epe$ is at the lower side of the experimentally
allowed range.

While such a combination of conditions on both the supersymmetric
models and the hadronic parameters is not very likely to be realized,
it cannot be rigorously excluded either. We conclude that models
that combine alignment and approximate CP are disfavored by the
measurement of $\epe$.

\newsec{Heavy Squarks}
Most of the supersymmetric flavor and CP problems are solved if the masses 
of the first and second generation squarks $\tilde m_h$
are larger than the other soft masses, $\tilde m_l$: $\tilde m_h^2\sim 
100\, \tilde m_l^2$
\nref\DKL{M. Dine, A. Kagan and R.G. Leigh,
 Phys. Rev. D48 (1993) 4269, hep-ph/9304299.}%
\nref\DiGi{S. Dimopoulos and G.F. Giudice,
 Phys. Lett. B357 (1995) 573, hep-ph/9507282.}%
\nref\PoTo{A. Pomarol and D. Tommasini,
 Nucl. Phys. B466 (1996) 3, hep-ph/9507462.}%
\refs{\DKL-\PoTo}. This does not necessarily lead to naturalness
problems, since these two generations are almost decoupled from the Higgs
sector. Explicit models are presented in
\nref\CKNS{A.G. Cohen, D.B. Kaplan and A.E. Nelson,
 Phys. Lett. B388 (1996) 588, hep-ph/9607394.}%
\nref\CKLN{A.G. Cohen, D.B. Kaplan, F. Lepeintre and A.E. Nelson, 
 Phys. Rev. Lett. 78 (1997) 2300, hep-ph/9610252.}%
\nref\DvaPo{G. Dvali and A. Pomarol,
 Phys. Rev. Lett. 77 (1996) 3728, hep-ph/9607383.}%
\nref\MoRi{R.N. Mohapatra and A. Riotto, 
 Phys. Rev. D55 (1997) 1138, hep-ph/9608441.}%
\nref\NeWr{A.E. Nelson and D. Wright,
 Phys. Rev. D56 (1997) 1598, hep-ph/9702359.}%
\nref\AmNe{S. Ambrosanio and A.E. Nelson, Phys. Lett. B411 (1997) 283, 
 hep-ph/9707242.}%
\nref\Okad{N. Okada, Prog. Theor. Phys. 99 (1998) 635, hep-ph/9711342.}%
\nref\ALT{N. Arkani-Hamed, M.A. Luty and J. Terning,
 Phys. Rev. D58 (1998) 015004, hep-ph/9712389.}%
\nref\AgGr{K. Agashe and M. Graesser,
 Phys. Rev. D59 (1999) 015007, hep-ph/9801446.}%
\nref\FKP{J.L. Feng, C. Kolda and N. Polonsky,
 Nucl. Phys. B546 (1999) 3, hep-ph/9810500.}%
\nref\BFP{J. Bagger, J.L. Feng and N. Polonsky, hep-ph/9905292.}%
\nref\KaKr{D.E. Kaplan and G.D. Kribs, hep-ph/9906341.}%
\refs{\PoTo,\CKNS-\KaKr}. We here follow mainly the discussion in
\refs{\CKNS-\CKLN}\ but our main points are of more general validity.

The main feature of these models that is relevant to our discussion
of $\epe$ is that the supersymmetric breaking scale that appears in
the $A$ terms is (at most) of the order of the electroweak scale 
while the scale that characterizes the average mass of the down and 
strange squark masses is $\tilde m_h$. Consequently, the $\delta_{LR}$ 
parameters are strongly suppressed:
\eqn\heavyLR{(\delta_{LR}^q)_{ij} \lsim {m_Z (m_q)_{ij} \over\tilde m_h^2} \sim 
10^{-4}{(m_q)_{ij}\over m_Z}.}
In this expression, $(m_q)_{ij}$ is related to the flavor structure
of the model, which is not always defined in the above models.
In any case, it is $\lsim m_b$ ($m_t$) in the down (up) sector.
Independent of the details of the model we have then 
\eqn\delud{\eqalign{
(\delta_{LR}^d)_{ij}\ \lsim&\ 10^{-4}(m_b/m_Z)\sim5\times10^{-6},\cr 
(\delta_{LR}^u)_{ij}\ \lsim&\ 10^{-4}(m_t/m_Z)\sim2\times10^{-4}.\cr}}
Comparing these upper bounds to eqs. \lowerLRot\ and \weakZds, we learn
that, similarly to the models of Abelian horizontal symmetries, only 
$(\delta_{LR}^d)_{12}$ and $(\delta_{LR}^d)_{21}$ can account for $\epe$.
  
The suppression from the large $\tilde d$ and $\tilde s$ masses is
not enough, however, to satisfy the $\Delta m_K$ constraint.
A mild alignment, $(K^d_L)_{12}\sim\lambda$, is required. Therefore,
the upper bound on $(\delta_{LR}^d)_{12}$ and $(\delta_{LR}^d)_{21}$
is actually of ${\cal O}(10^{-6})$. If we make the further plausible
assumption that this alignment reflects a flavor structure that
is similar to that of the corresponding quark mass matrix, we have
\eqn\HSali{(\delta_{LR}^d)_{12}\ \lsim\ 10^{-4}{m_s|V_{us}|\over m_Z}
\sim3\times10^{-8}.} 
We learn that in the likely situation that $(m_d)_{12}\sim m_s|V_{us}|$,
$(\delta_{LR}^d)_{12}$ is already below the value where it could potentially
give a significant contribution to $\epe$. However, if the flavor structure
is such that $(m_d)_{12}\sim m_b|V_{us}|$, a significant contribution is
not excluded. We emphasize, however, that there is no flavor model that
predicts such a large value.

The combination of heavy squark masses and alignment of order of the
Cabibbo angle is still not enough to satisfy the $\epsK$ constraint. 
This constraint is satisfied if the CP violating supersymmetric phases 
in the down and strange couplings are less than ${\cal O}(1/30)$ \CKLN. 
Then, very likely, this phase also suppresses $\Im[(\delta^d_{LR})_{12}]$, 
that is,
\eqn\HSaliCP{\Im[(\delta_{LR}^d)_{12}]\ \lsim\ {m_Z m_s|V_{us}|\phi_{\rm CP}
\over\tilde m_h^2}\sim10^{-9}.} 
With a different flavor structure, we can imagine an enhancement of
this contribution by a factor of order, at most, $m_b/m_s$, that is
to $3\times10^{-8}$. In any case, this contribution is smaller than
the lower bound in \lowerLRot\ and, therefore, cannot account for $\epe$.
 
If we make the final assumption, that the smallness of CP violating
phases in the down and strange squark sector is not accidental and
special to this sector but rather reflects an approximate CP symmetry
in all interactions, that is
all CP violating phases are small, then the Standard Model contributions
are also too small and this class of models is excluded. We would like
to emphasize, however, the following two points:
\item{(i)} The motivation for approximate CP is weaker with heavy
squarks than it is with alignment. The reason is that the heavy squarks
suppress to a satisfactory level the supersymmetric contributions to 
electric dipole moments even for (flavor-diagonal) CP violating phases
of ${\cal O}(1)$. In contrast, alignment has no effect on flavor-diagonal
CP violation.
\item{(ii)} The $\epsK$ constraint applies to 
$\Im\{[(\delta^d_{LL})_{12}]^2\}$. It could be satisfied, therefore, not
only by a very small phase (that is, $\Im[(\delta^d_{LL})_{12}]\ll|
(\delta^d_{LL})_{12}|$) but also by a phase that is very close to
$\pi/2$ (that is, $\Re[(\delta^d_{LL})_{12}]\ll|(\delta^d_{LL})_{12}|$).
Then our discussion here of approximate CP is irrelevant. We note, however,
that we know of no model which predicts such a near-maximal phase.

\newsec{Conclusions}
Supersymmetric models suffer from two problems related to CP violation.
First, the $\phi_A$ and $\phi_B$ phases give an electric dipole moment
of a neutron that is two orders of magnitude above the experimental bound,
unless the phases are small or squarks of the first two generations heavy.
Second, in models without universality, phases in the mixing matrices
for gaugino couplings to quarks and squarks give a value to $\epsK$
that is higher than the experimental value. There are three known ways
in this type of models to suppress these contributions:
\item{(i)} Horizontal non-Abelian symmetries lead to approximate
degeneracy between the first two squark generations;
\item{(ii)} Horizontal Abelian symmetries lead to suppression of
the mixing angles by alignment of the mass matrices;
\item{(iii)} The first two squark generations could be heavy.

The models with horizontal (Abelian or non-Abelian) symmetries do
not solve in general the $d_N$ problem. 
Whether the $\epsK$ problem is solved depends on the details of the
model. If, in addition to employing one of these mechanisms to
suppress flavor violation, there is also approximate CP to suppress
CP violation, then both the $d_N$ problem and the $\epsK$ problem
are solved.

The models with heavy squarks do solve the $d_N$ problem but, in general,
neither the $\Delta m_K$ nor the $\epsK$ constraints are satisfied.
Alignment of order of the Cabibbo angle can solve the first and small 
CP violating phases the second problem.

If there is an approximate CP in nature, then the Standard Model
with $\delta_{\rm KM}\ll1$ cannot account for the measured value of $\epe$.
If the supersymmetric models with approximate CP are to be viable 
then they have to provide a supersymmetric mechanism to explain $\epe$. 

We have first examined this problem in the framework of alignment
and reached the following conclusions:
\item{(i)} Models that combine alignment (to solve
the supersymmetric flavor problem) and approximate CP (to solve the
supersymmetric CP problems) are strongly disfavored. Only if several
model parameters as well as several hadronic parameters take rather
extreme values, the models are viable. 
\item{(ii)} Models of Abelian horizontal symmetries and approximate CP
where the flavor problems are solved by a mechanism different from alignment 
can account for $\epe$.
\item{(iii)} Models of alignment without approximate CP are likely to
give a significant supersymmetric contribution to $\epe$, in addition to
the Standard Model contribution.

We then examined the problem in the framework of heavy squarks
and reached the following conclusions:
\item{(i)} Models that combine heavy first two squark generations
with alignment of order of the Cabibbo angle to solve the supersymmetric 
flavor problem and employ approximate CP to solve the
supersymmetric $\epsK$ problem are excluded. 
\item{(ii)} Models of heavy squarks without approximate CP are still
unlikely to give a significant supersymmetric contribution to $\epe$.

In the near future, we expect first measurements of various CP asymmetries 
in $B$ decays, such as $B\rightarrow\psi K_S$ or $B^\pm\rightarrow\pi^0 K^\pm$.
If these asymmetries are measured to be of order one, it will support the 
Standard Model picture, that the CP violation that has been measured in the 
neutral $K$ decays is small because it is screened by small mixing angles, 
while the idea that CP violation is small because all CP violating
phases are small will be excluded. It is interesting, however, that various 
specific models that realize the latter idea, such as those discussed in this 
work, can already be excluded by the measurement of a tiny CP violating effect,
$\epspK\sim5\times10^{-6}$.

After the completion of this work, a new lattice result appeared
\ref\Blum{T. Blum {\it et al.}, hep-lat/9908025.}. It finds that the value
of the hadronic matrix element that is relevant to the standard model
contribution to $\epe$ is larger by about one order of magnitude than (and
of opposite sign to) its value in the vacuum insertion approximation. If
this surprising result is confirmed, then the framework of supersymmetry
with alignment and with approximate CP becomes attractive: $\epsK$ is
naturally accounted for by supersymmetric contributions with a small phase
in $(\delta^d_{LL})_{12}(\delta^d_{RR})_{12}$ while $\epe$ is naturally 
accounted for by the standard model contributions with a small 
$\delta_{\rm KM}$ 
\ref\prog{G. Eyal {\it et al.}, work in progress.}.
In particular, model I of ref. \Eyal\ is a viable example of this idea.

\vskip 0.6 cm
\centerline{\bf Acknowledgements}
The work of A.M. is partially supported by the EEC TMR Network ``BSM" 
under contract ERBFMRX CT960090.
Y.N. is supported in part by the United States $-$ Israel Binational
Science Foundation (BSF), by the Israel Science Foundation founded
by the Israel Academy of Sciences and Humanities
and by the Minerva Foundation (Munich).
L.S. acknowledges the support of the German Bundesministerium f{\"u}r
Bildung und Forschung under contracts 06 TM 874 and 05 HT9WOA.

\listrefs
\end